\documentclass[aps,prb,twocolumn,preprintnumbers,amsmath,amssymb]{revtex4}

\usepackage{graphicx}
\usepackage{dcolumn}
\usepackage{bm}
\usepackage{textcomp}
\usepackage{rotating,booktabs}
\begin{document}
\input epsf.sty

\title{Instrumentation of a high-sensitivity microwave vector detection system for low-temperature applications}

\author{Y.\ W.\ Suen,$^{\textrm{a)}}$
\footnotetext{$^{\textrm{a)}}$Author to whom correspondence should
be addressed; electronic mail:ysuen@phys.nchu.edu.tw} W.\ H.\
Hsieh, C.\ L.\ Chen, and L.\ C.\ Li}

\affiliation{ Department of Physics, National Chung Hsing
University, No. 250, Kuo-Kuang Rd., Taichung 402, Taiwan, Republic
of China}

\author{C.\ H.\ Kuan}

\affiliation{ Department of Electrical Engineering, and Graduate
Institute of Electronics Engineering, National Taiwan University,
Taipei, Taiwan, Republic of China}

\date{\today}
\begin{abstract}
We present the design and the circuit details of a
high-sensitivity microwave vector detection system, which is
aiming for studying the low-dimensional electron system embedded
in the slots of a coplanar waveguide at low temperatures. The
coplanar waveguide sample is placed inside a phase-locked loop;
the phase change of the sample may cause a corresponding change in
the operation frequency, which can be measured precisely. We also
employ a double-pulse modulation on the microwave signals, which
comprises a fast pulse modulation for gated averaging and a slow
pulse modulation for lock-in detection. In measurements on real
samples at low temperatures, this system provides much better
resolutions in both amplitude and phase than most of the
conventional vector analyzers at power levels below -65 dBm.

\end{abstract}

\maketitle

\section{Introduction}

In the study of dynamic magnetotransport behaviors of
low-dimensional electron systems (LDESs), such as two-dimensional
electron systems
(2DESs),\cite{Engel1993,Li1997,Ye2002,Lewis2002,Chen2003} quantum
wires (QWs)\cite{Grodnensky1994} , and anti quantum dots
(QDs),\cite{Ye2002B}  a microwave vector detection system, which
can measure both the amplitude and the phase variations of the
signals under the influence of the LDESs in magnetic fields, is
indispensable. Usually people use a conventional vector network
analyzer (VNA) as such a detection system by virtues of its
expediency and broadband characteristics. The LDES under test can
be either embedded in the slots of a coplanar waveguide
(CPW),\cite{Engel1993,Li1997,Ye2002,Lewis2002,Chen2003,Ye2002B}
attached to a resonator,\cite{Lewis2001} or directly connected to
a coaxial cable;\cite{Hohls2001} then the transmission or
reflection property can be measured, and the transport properties
of the LDES can be extracted therefrom, as the external magnetic
field ($B$), or the temperature ($T$) changes. However, since in
most of these studies the LDES is placed in a very cold
environment with temperature ($T$) below 4.2 K, or even down to
several tens of minikelvins, the power of the excitation signal
must be very low, normally below $-$60 dBm, to avoid joule
heating. Moreover, the power reaching the detection system is even
lower due to the loss of coaxial cables and the sample itself.
Hence the resolution of the data, in particular the phase part,
becomes very poor.

In this article, we present the instrumentation and the details of
circuit design of a very-high-sensitivity vector detection
system,\cite{Hsieh2004} incorporating phase-locked loop (PLL) and
double pulse techniques.  Here we mainly emphasize on the
application of this system to the study of LDESs embedded in the
slots of a CPW, which  have been successfully used as broadband
sensors.\cite{Engel1993,Li1997,Ye2002,Lewis2002,Chen2003,Ye2002B}
Our method is especially useful for the case of QDs or QWs, whose
relatively small effective area compared to 2DES samples usually
results in a very small variation on the transmission coefficient
of the CPW sample, and makes the conventional VNA measurement very
formidable and impractical.

In a typical modern VNA, a PLL is only used in the very front end
of its superheterodyne receiver\cite{White1993} to make the local
oscillator (LO) of the first-stage mixers follow the reference
signal from the tracking source, and thus to preserve the phase
information in the detected signal. In contrast, we put the CPW
sample under test in the high-frequency signal path of a PLL,
which is renowned for its phase sensitivity, and obtain a great
improvement in the phase resolution. As a matter of fact, the
PLL-based phase detection technique has been applied in the study
of 2DESs using surface acoustic wave (SAW).\cite{Wixforth1989} We
improve the design of their measurement system, including a better
pulse averaging part, a homodyne amplitude detection part and a
loop filter of the PLL, and also replace the SAW transducers with
a coaxial cable delay line and a broadband CPW containing a LDES
under test. All these result in a very sensitive microwave vector
detection system, which can be operated with extremely low power
levels through samples, making this system very suitable for
low-temperature applications.

This system has been used to study the dynamic magnetotransport
properties of a QW-array embedded in a CPW\cite{Hsieh2004} with an
average power below -65 dBm into the sample at 0.3 K, and the
resolution of the data, including both the amplitude and phase
variations, surpassed what most of the commercial VNAs can get at
such a low power level. We have also used this method to obtain
very clean data of the high-frequency longitudinal conductivity,
including both the real and the imaginary parts, of a 2DES in the
quantum Hall plateau.

\section{System and Circuit Designs}

\subsection{Phase Measurement with a PLL}

The principle of
 phase detection by a PLL\cite{PLL} is quite straightforward. Two semirigid coaxial cables of total
length $L$ connect the PLL and the CPW sample. The PLL can tune
its operation frequency ($f$) to ensure the sum ($\Delta \phi$) of
the phase change of the semirigid cables ($\Delta \phi_{L}$) and
the CPW sample ($\Delta \phi_{s}$) to be 0, i.e. $\Delta \phi =
\Delta \phi_{L}+\Delta \phi_{s}=0$, or $\Delta \phi_{s}=-\Delta
\phi_{L}$. Here $\Delta \phi_{L}$ can be easily obtained from
$\Delta \omega\tau_{L}$ or $2\pi \Delta fL/v_{L}$, where $\omega$
is the angular frequency ($2\pi f$), $v_{L}$  the phase velocity
of the signal in the cable, and $\tau_{L}$ the delay time of the
coaxial cable. Thereby we can obtain $\Delta \phi_{s}$ directly
from measuring the frequency change ($\Delta f$) of the PLL via
\begin{eqnarray}
\Delta \phi_{s}=-\Delta \phi_{L}=-2\pi \Delta fL/v_{L}=-\Delta
\omega\tau_{L}.
\end{eqnarray}

\subsection{Description of the Complete System }

\begin{figure}[h]
\includegraphics[width=8cm]{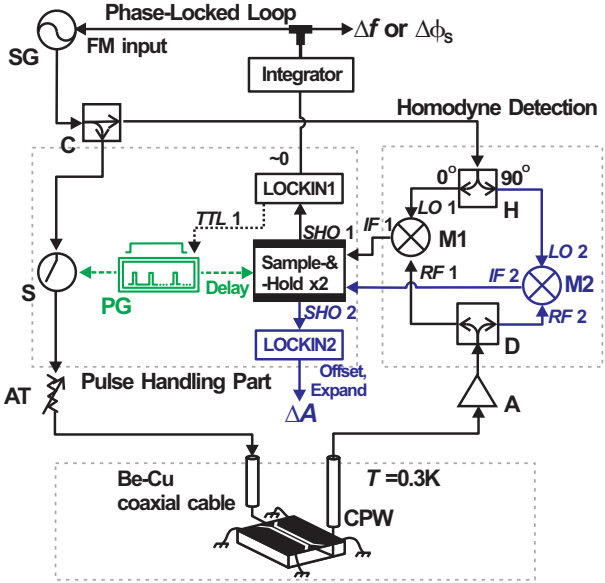}\\
\caption{(Color online) Simplified schematic diagram of the vector
detection system. The meandering CPW containing a LDES in the gaps
is part of the microwave signal path in the PLL.}
\end{figure}

Figure 1 shows the complete schematic diagram of the detection
system, consisting of the pulse handling part, the microwave PLL,
and the amplitude readout part, together with the CPW sample in a
cryogenic environment. The main microwave components of this
system are basically a pair of homodyne mixers (M1 and M2) with
different reference signals of quadrature phase difference
generated by a 90\textdegree\ hybrid (H). The signal from the
sample is divided  equally into two part by a power divider (D).
The mixer M1 with a 0\textdegree\ reference (\textit{LO}1), used
as the phase sensitive detector (PSD), has zero output
(\textit{IF}1) forced by the PLL, and at the same time the other
mixer M2 with a 90\textdegree\ reference (\textit{LO}2) has an
output (\textit{IF}2) proportional to the amplitude of the signal.

\begin{figure}[h]
\includegraphics[width=8cm]{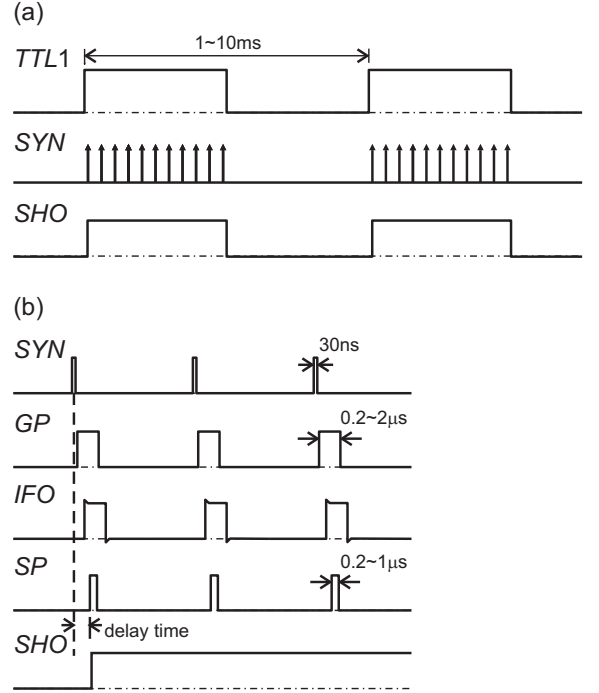}\\
\caption{Time sequence of the controlling, sampling, and some
output pulses. The time scale in (b) is much shorter than (a).}
\end{figure}

 To reduce the signal power level and increase the signal to noise
ratio, besides homodyne detection we also employ a double-pulse
modulation scheme to detect and average the microwave signal.
Figure 2 shows the sequence of the controlling pulses and some
other related outputs.
 A short pulse train [\textit{GP} in Fig. 2(b)] with a 0.2$\sim$2 $\mu$s pulse width and a
0.1$\sim$10\% duty cycle, provided by a pulse generator (PG in
Fig.1) and gated by a slow square-wave TTL signal (\textit{TTL}1)
shown in Fig. 2(a) with a period of 1$\sim$10 ms from a lock-in
amplifier, modulates the microwave signal sent to the sample. A
time-delayed pulse (\textit{SP}) with a 0.2$\sim$1 $\mu$s pulse
width, triggered by the modulating pulses, controls a
sample-and-hold (S\&H) circuit that samples the
 IF output (\textit{IFO}) of the microwave mixer. The output of
 the S\&H circuit (\textit{SHO}) is set to zero by an analog
switch  when the TTL gating signal is low. The output of the S\&H
circuit has the same characteristic frequency as the signal
\textit{TTL}1, and hence can be averaged and read out by the
lock-in amplifier.

We use two sets of pulse averaging circuits in this system, one
for the PLL part and the other for the amplitude detection part.
Even though the lock-in amplifiers therein are synchronized at the
same operating frequency, their averaging time constants are
different. The time constant of LOCKIN1 in the PLL is about
1$\sim$10 ms for fast loop dynamics, in contrast to 300 ms or 1 s
of LOCKIN2 in the amplitude readout part for low average noise.
The average of the PSD output (\textit{IF}1) is then sent to an
integrator (loop filter of PLL), whose output is connected to the
frequency modulation (FM) input of the microwave source (SG), i.e.
the VCO of the PLL, and thus the loop is closed.

\subsection{Pulse Handling Circuits and the Integrator}

The short pulse trains used in this system are provided by a
simple single-channel pulse generator TPG110 from TTi\cite{TTi},
which can generate a short reference pulse train (\textit{SYN} in
Fig. 2) and a pulse train (\textit{SP}) with tunable time delay
(relative to \textit{SYN}), pulse width, and duty cycle, and a
homemade pulse shaping circuit (Fig. 3), which converts the
30-ns-width reference pulses (\textit{SYN}) to the gating pulses
(\textit{GP}) with 0.2$\sim$2 $\mu$s pulse width that are sent to
a diode switch (S) to modulate the microwave signal. A lock-in
amplifier (LOCKIN1), SR830 from SRS,\cite{lockin} sends a slow TTL
square-wave reference (\textit{TTL}1) to the "trigger/gate in"
input of TPG110. Another lock-in amplifier (LOCKIN2) is externally
locked to LOCKIN1 and used to read the IF output of M2
(\textit{IF}2).

\begin{figure}[h]
\includegraphics[width=8cm]{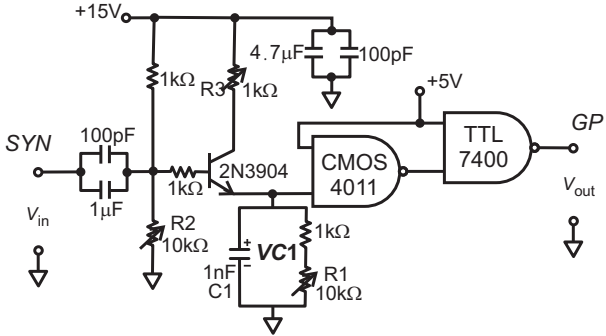}\\
\caption{Simple pulse shaping circuit to generate a 0.2-2 $\mu$s
pulse from a 30-ns input pulse.}
\end{figure}

The simple pulse shaping circuit in Fig. 3 consists of a BJT
switch (2N3904), a CMOS and a TTL invertors. During the 30-ns
duration given by \textit{SYN}  pulses, 2N3904 is turned on and
allows certain amount of charge (set by R2 and R3) to flow into C1
and the TTL output (\textit{GP}) will be set to high from low.
Afterwards the voltage across C1 (\textit{VC}1) drops below 2.5 V
due to the discharging current through R1 (in series with a 1
k$\Omega$ resistor), that triggering the following invertors to
switch their statuses and thus making the output (\textit{GP})
back to low. The width of the output pulse can be tuned via the
variable resistor R1, and the tunable range of the pulse width can
be adjusted by R2 and R3.

\begin{figure}[h]
\includegraphics[width=8cm]{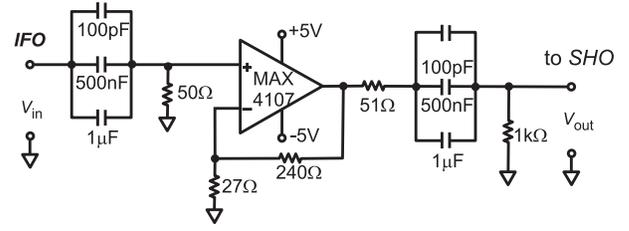}\\
\caption{Circuit of a pulse amplifier.}
\end{figure}

Each IF output of the mixers (M1 and M2) is low-pass filtered and
then amplified by a pulse amplifier before fed into the
sample-and-hold circuit. The low-pass filters used here are PLP-5
($f_{3\textrm{dB}}=$5MHz) or PLP-70 ($f_{3\textrm{dB}}=$70MHz)
from Mini-Circuits,\cite{minicircuits} depending on the pulse
width of the gating pulse (\textit{GP}). For the pulse amplifier,
we use a high-speed low-noise operational amplifier
MAX4107\cite{maxim} connected in a non-inverted configuration as
shown in Fig. 4. Its rise/fall time is well below 10 ns for a
typical pulse height ($\lesssim$1 V) in our application.

\begin{figure}[h]
\includegraphics[width=8cm]{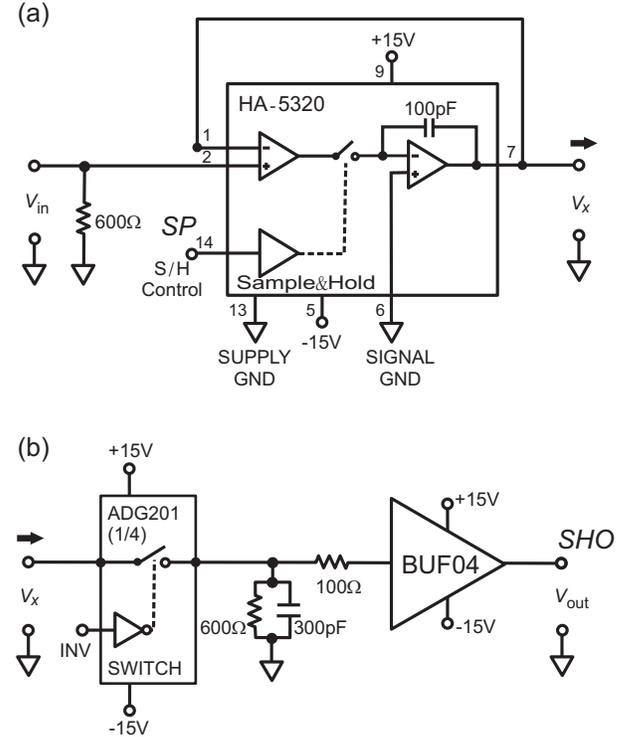}\\
\caption{(a) Sample-and-hold circuit with (b) an analog switch and
a buffered output.}
\end{figure}

The sample-and-hold circuit is shown in Fig. 5. We use a
high-speed sample-and-hold amplifier IC HA-5320-5,\cite{intersil}
which provides a 100 ns rise time and a 25 ns aperture time, and a
fast analog switch ADG201HS\cite{ADI} followed by a buffer
BUF04\cite{ADI}. The main output of TPG110 (\textit{SP}) is
connected to the S/H control (lead 14) of HA-5320-5, and the slow
reference \textit{TTL}1 from LOCKIN1 controls ADG201HS and makes
the output \textit{SHO} zero as \textit{TTL}1 goes
low.\cite{discharge}

\begin{figure}[h]
\includegraphics[width=8cm]{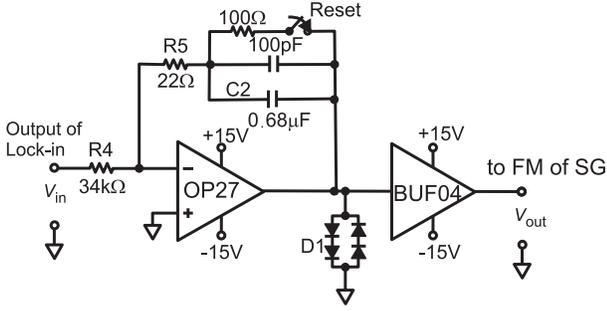}\\
\caption{Circuit of the integrator (loop filter) in the phase-lock
loop.}
\end{figure}

A simple integrator shown in Fig. 6 serves as the loop filter of
the PLL. The integration time constant ($R_{4}C_{2}$) is about 23
ms, which is chosen to be higher then the time constant of LOCKIN1
such that the dynamics of the PLL is mainly controlled by this
integrator. Two sets of series-connected diodes (D1) are connected
in parallel from the output of the integrator to the ground with
inverted polarities and used as a voltage limiter to protect the
FM input of the microwave generator (SG). A manual switch is used
to reset the integrator while the PLL is saturated or unlocked. R5
gives the PLL an adequate damping that may attenuate unwanted
oscillations in the loop. The dynamic properties of this PLL
system will be discussed later in Sec. III.

\subsection{Microwave Modules}

We use several sets of microwave modules to cover the frequency
range from about 100 MHz to 18 GHz. Table I lists the model
numbers of the modules (including the vendors in the
footnotes\cite{vendors}) according to their operation band. Please
note that the amplifers (A) includes a front-end low-noise
preamplifier (LNA) and subsequent amplifiers to make the total
gain about 60 dB. To cover the whole band, we use two LNAs with
noise figure (NF) better then 1.4 dB and 2.2 dB for frequency
range 0.1$\sim$8 GHz and 6$\sim$18 GHz, respectively. The
microwave signal source (SG) used here is model E4421B (250k-3GHz)
or E8254A (250k-40GHz) from Agilent,\cite{vendors} both with FM
bandwidth down to DC.

\begin{table*}
\renewcommand{\arraystretch}{1.5}
\caption{The model numbers of the microwave
modules.}\label{tab:modules} \vspace{0.5cm}
\begin{tabular}{|c|c|c|c|c|c|c|}\hline

 {\renewcommand{\arraystretch}{1.2}\begin{tabular} {c}Band\\(GHz)\end{tabular}} &
  Diode
Switch (S)& {\renewcommand{\arraystretch}{1.2}\begin{tabular} {c}
Mixer\\\hline(M1\& M2)\end{tabular}} &
 90\textdegree \ Hybrid (H)&
 Divider  (O) &
 LNA + Amp (A) &
{\renewcommand{\arraystretch}{1.2}\begin{tabular}{c} Directional\\
Coupler
 (C)\end{tabular}}\\\hline\hline
{\renewcommand{\arraystretch}{1.2}\begin{tabular}{c}0.1-0.5\\\hline
0.4-2 \end{tabular}} & ZASWA-2-50DR$^{a}$ & ZFM-2000$^{a}$ &
{\renewcommand{\arraystretch}{1.2}\begin{tabular}{c}QE-18-E$^{c}$\\\hline
QS4-01-464/6$^{c}$
\end{tabular}}& ZFSC-2-2500$^{a}$ &
{\renewcommand{\arraystretch}{1.2}\begin{tabular}{c}
AFS3-00100800-14-10P-4$^{b}$\\+ ZJL-7G$^{a}\times 3$
\end{tabular}} &
{\renewcommand{\arraystretch}{1.2}\begin{tabular}{c}ZFDC-20-5$^{a}$\\(-19.5$\pm$0.5dB)\end{tabular}}
\\\hline

\ 2-4.3&
 ZASWA-2-50DR$^{a}$ &
  ZEM-4300$^{a}$ &
QS4-01-464/3$^{c}$ &
 ZFSC-2-10G$^{a}$ &
{\renewcommand{\arraystretch}{1.2}\begin{tabular}{c}
AFS3-00100800-14-10P-4$^{b}$\\ + ZJL-7G$^{a}\times 3$
\end{tabular}} &
{\renewcommand{\arraystretch}{1.2}\begin{tabular}{c}86205A$^{f}$
\\ (-16.5dB) \end{tabular}}
\\\hline

{\renewcommand{\arraystretch}{1.2}\begin{tabular}{c}4-7\\\hline
6-18\end{tabular}} & N138BDF1$^{b}$ & DB0218LW2$^{b}$ &
QS2-07-464/4$^{c}$ &
 2089-6209-00$^{d}$&
{\renewcommand{\arraystretch}{1.2}\begin{tabular}{c}
AFS3-00100800-14-10P-4$^{b}$\\ + ZJL-7G$^{a}\times 3$\\\hline
AFS5-06001800-22-10P-6$^{b}$\\ + QLW-12124033$^{d}$ \end{tabular}}
& {\renewcommand{\arraystretch}{1.2}\begin{tabular}{c}
87300B$^{f}$\\(-16dB) \end{tabular}}\\\hline

\end{tabular}
\\\vspace{0.5cm} $^{a}$ Mini-circuits; $^{b}$ MITEQ; $^{c}$ Quinstar;
$^{d}$ M/A COM; $^{e}$ Pulsar Microwave; $^{f}$ Agilent.
\end{table*}

\section{Results and Discussions}

\subsection{Response of the PLL}

The response of this PLL system is mainly controlled by four
parameters, the power of the signal from the sample back to the
LNA, the sensitivity of LOCKIN1 ($V_{sen}$), the FM deviation per
volt ($F_{FM}$, with unit Hz/V) of the VCO, and the total length
of the semirigid coaxial cables, all of which can be tuned by the
instrumental setting in our design. The length of the semirigid
cables affects the delay time $\tau_{L}$ in Eq. 1. We can measure
$\tau_{L}$ precisely with this PLL system itself without using an
additional conventional VNA or time domain reflectometer (TDR). In
the unlock condition, i.e. with FM off in SG (VCO), we recorded
the output of LOCKIN1 (or LOCKIN2) as a function of frequency $f$,
and the result can be expressed as $A_{ul}\exp(2\pi f
\tau_{L}+\varphi_{0})$ with $A_{ul}$ the oscillation amplitude,
and $\varphi_{0}$ a relative phase factor. Figure 7(a) shows the
oscillations for both LOCKIN1 and LOCKIN2 near $f$ = 1.2 GHz with
only a 10.4-m semirigid cable and no sample connected to the PLL
system between the attenuator (AT) and A in Fig.1. As expected,
the oscillations of these two channels are 90\textdegree\ out of
phase due to the LO phase difference of two mixers. From the
``frequency'' of these oscillations, we can directly estimate
$\tau_{L}$ to be 53.3 ns. Figure 7(b) shows that $\tau_{L}$ is a
linear function of the length of the semirigid cable with a finite
intercept, about 2.6 ns, on the vertical axis. This extra delay
time corresponding to a zero-length cable can be attributed to the
time delay in the microwave modules. The peak power of the
microwave signals used in these tests is -70 dBm.

\begin{figure}[h]
\includegraphics[width=8cm]{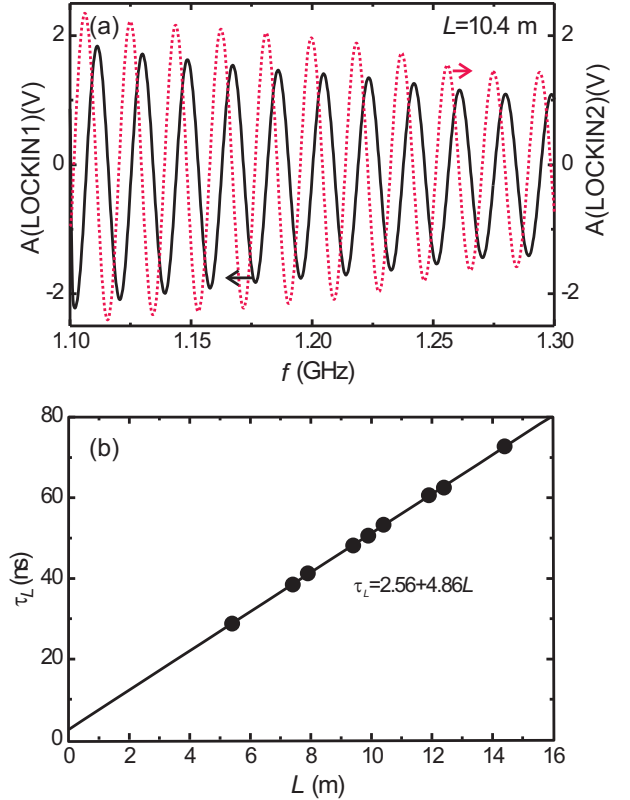}\\
\caption{(Color online) (a) Output oscillations of LOCKIN1 and
LOCKIN2 with only a 10.4-m semirigid cable and no sample connected
to the PLL system. (b) Delay time $\tau_{L}$ versus length ($L$)
of the semirigid cable.}
\end{figure}

The phase change of the CPW device in the experiment is the
product of the output of the integrator $V_{FM}$ in the PLL,
$F_{FM}$ and $\tau_{L}$ while keeping the PLL in the locked
condition. $V_{FM}F_{FM}$ is in fact the change in the frequency
($\Delta f$) of the VCO, which can also be measured by a microwave
counter. $F_{FM}$ of the signal source may affect the phase
resolution: the smaller is the FM deviation set, the higher is the
phase sensitivity. However, if $F_{FM}$ is set too small, the PLL
may saturate very easily (the output of the integrator $V_{FM}$
reaches $+1.4$ or $-1.4$ V), and the phase signal readout $V_{FM}$
is also very noisy. Usually $F_{FM}$ needs to be set properly
according to the dynamic range of the phase signal.

 For low-temperature experiments, usually the signal power to the sample is set by a
step attenuator and is chosen to be the highest value without any
heating effect on the sample. For samples immersed in liquid He3
at 0.3 K, typically a peak power of -50 dBm microwave signal,
equivalent to an average power of -65 to -80 dBm depending on the
duty cycle of the modulating pulses, can be delivered to the
sample without raising the sample temperature.

The response time of the PLL is affected by the total gain of the
loop, which can be obtained from the amplitude of the oscillations
in Fig. 7(a) and $F_{FM}$. $A_{ul}$ can be easily controlled by
$V_{sen}$ of LOCKIN1, and this provides a very convenient method
to control the response time of the PLL during the experiment.
Figure 8 shows the step response for different value of $V_{sen}$.
In this test, we also used a 10.4-m semirigid cable without sample
attached to the PLL system. $F_{FM}$ was set at 1 MHz/V, the peak
power of the microwave signal after the attenuator was -70 dBm,
and the operation frequency was about 1.2 GHz. The step response
was measured from the output of the integrator $V_{FM}$ as
function of time right before we increased the unmodulated
frequency of SG by a step of 500 kHz while the PLL was kept in a
locked condition. The rise time of the response can be tuned from
0.63 s for $V_{sen}=1$ V to 0.015 s for $V_{sen}=50$ mV, and at
the same time the loop dynamic behavior is moved from an
overdamped regime to a slightly underdamped regime. We also find
that the fluctuation of the signal decreases as $V_{sen}$
increases. There is apparently a tradeoff between the speed of the
loop and the signal fluctuation.

\subsection{Detection Limit}
To gain an idea of the detection limit of this system, we measured
the background noise in the phase signal with only a 5.4-m
semirigid cable attached to the system. Various microwave modules
were used to cover the whole available frequency range. Figure 9
shows the measured peak-to-peak phase fluctuation versus frequency
for different peak power levels of the microwave signal after the
step attenuator (AT) together with the loss of the semirigid cable
measured by a VNA. The phase signal was read from $V_{FM}$ and
then multiplied by $F_{FM}\tau_{L}$.
 The measured bandwidth is controlled by the time constant of the integrator (23 ms), equivalent to about 7 Hz.
   $V_{sen}$ was fixed at 500 mV and $V_{FM}$
was set at 50 kHz/V during this test. The modulating pulses had a
2 $\mu$s pulse width and a 1/75 duty cycle. The gating window of
the S\&H circuit was 350 ns.

\begin{figure}[h]
\includegraphics[width=8cm]{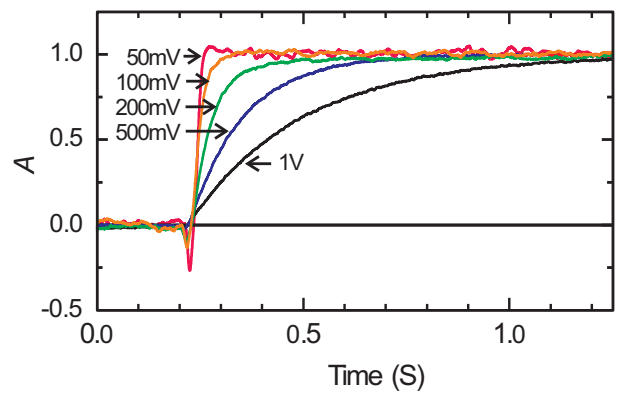}\\
\caption{(Color online) Normalized step responses measured from
the output of the integrator with different sensitivity levels of
LOCKIN1.}
\end{figure}

The collected data shown in Fig. 9 indicate that below 5 GHz, even
for the lowest peak power -90 dBm in our test, the phase
fluctuation is still less than 0.0003\textdegree; above 5 GHz to
18 GHz, the fluctuation is still controlled within about
0.002\textdegree. This noise level is remarkably low for such a
low-power signal. Here we want to note that to obtain the average
power value we must further subtract about -21.8 dBm from the peak
power value due to the low duty cycle. In fact, the signal power
reaching the low-noise amplifier (LNA) is even lower than the
input value claimed above due to the loss of the cable as also
shown in Fig. 9. This may explain the noise in phase increases at
high frequencies. The resolution with a low-$T$ sample loaded is
slightly worse due to the loss of the sample and extra noise from
the cryogenic environment. The resolution of the amplitude readout
for a small-variation signal can be enhanced by the use of the
"offset" and "expand" functions of the lock-in
amplifier.\cite{lockin}

\begin{figure}[h]
\includegraphics[width=8cm]{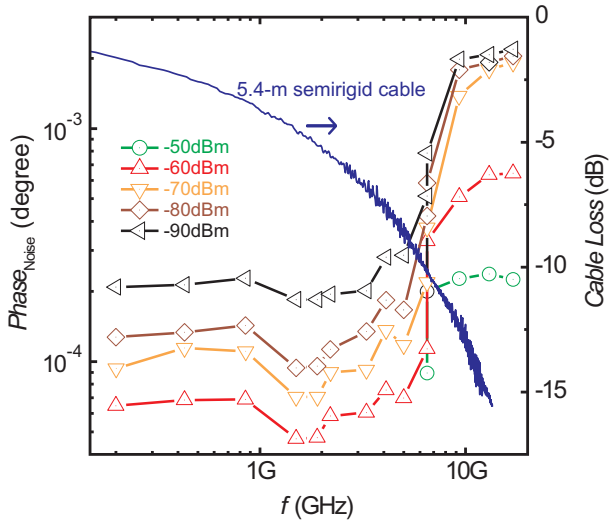}\\
\caption{(Color online) Background peak-to-peak noises in the
phase signal as functions of frequency $f$ tested with different
power levels. The loss of a 5.4-m semirigid cable is also shown
for comparison.}
\end{figure}

\subsection{Applications to Low-Dimensional Electron Systems}

In the following we will present measured results for a 2DES
sample (sample A) and a QW array sample (sample B) to demonstrate
the capability of this system. The 2DES in the sample A is
confined in a 20-nm-wide modulation-doped In$_{0.15}$Ga$_{0.85}$As
quantum well grown on GaAs substrate by molecular beam epitaxy
(MBE) technique. The quantum well is separated from a 260 nm
Al$_{0.3}$Ga$0.7$As Si-doped layer by a 15-nm GaAs and a 5nm
Al$_{0.3}$Ga$0.7$As undoped spacer layers. This 2DES has an
electron density of $5.95\times10^{11}$ cm$^{-2}$ and a mobility
of about $3.2\times10^{5}$ cm$^{2}/$V s at 0.3 K. A 50-$\Omega$
($Z_{o}$) meandering CPW made of Au(300 nm)/Ti(10 nm) with a 36
$\mu$m wide central conductor and a 23 $\mu$m gap ($d$) between
the central conductor and the ground plane at both sides is
manufactured on the surface of the substrate. The effective length
$\ell$ of the meandering CPW is 2.25 cm. The sample is immersed in
liquid $^3$He (0.3 K) with applied $B$ perpendicular to the sample
surface. The real part and the imaginary part of the longitudinal
conductivity can be deduced from the relative amplitude
change\cite{Engel1993} of the microwave signal ($\Delta A/A$) and
the phase difference\cite{Suen2002} obtained from Eq. 1,
respectively, via

\begin{eqnarray}
\textrm{Re}\{\sigma_{xx}\}=d\cdot\frac{\ln|\Delta
A/A|}{Z_{o}\ell},
\end{eqnarray}
and
\begin{eqnarray}
\textrm{Im}\{\sigma_{xx}\}=d\cdot\frac{\Delta\phi_{s}}{Z_{o}\ell}.
\end{eqnarray}

Figure 10 shows the Re$\{\sigma_{xx}\}$ and Im$\{\sigma_{xx}\}$ of
the sample A at 6.99 GHz for $B$ between 3 and 10 T. The scale bar
on the plot is 10 $\mu$S for both curves. The peak power into the
sample was about -85 dBm, and the equivalence average power was
only about -105 dBm. The Re$\{\sigma_{xx}\}$ curve shows clear
broad minima near zero conductance around integer Landau level
filling factors ($\nu$), while the Im$\{\sigma_{xx}\}$ curve
exhibits a concave shape in the plateau regime and a deep minimum
in the transition region between adjacent Hall states. This
observation is very similar to what Hohls et al.\cite{Hohls2001}
reported; however, compared to their presented data (Fig. 2 in
Ref. 9), our data give a much less background noise than theirs.
The background fluctuation for both curves in Fig. 10 is only
about 50 nS.

\begin{figure}[h]
\includegraphics[width=8cm]{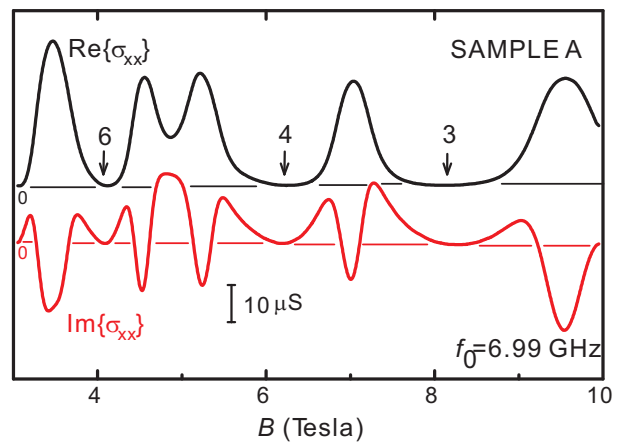}\\
\caption{(Color online) Measured Re$\{\sigma_{xx}\}$ and
Im$\{\sigma_{xx}\}$ of a 2DES in an InGaAs quantum well embedded
in the slots of a CPW at 0.3 K with the PLL operating at 6.99 GHz.
The integer number with an arrow marks the Landau level filling
factor at that field.}
\end{figure}

\begin{figure}[h]
\includegraphics[width=8cm]{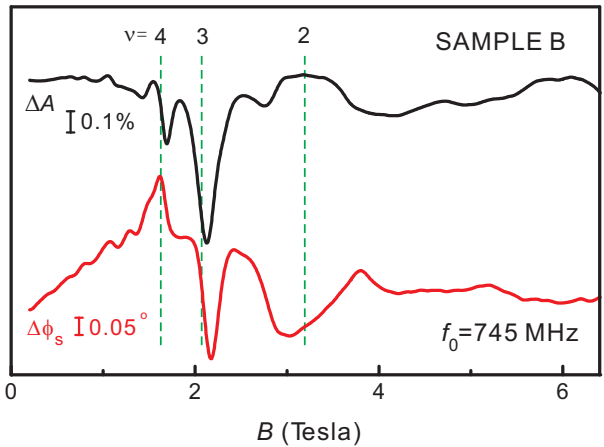}\\
\caption{(Color online) Relative amplitude change ($\Delta A$)and
the phase variation ($\Delta\phi_{s}$) of a CPW containing a
QW-array in the slots.}
\end{figure}

 The sample B is fabricated from a typical MBE-grown
modulation-doped GaAs/AlGaAs heterostructure containing a 2DES,
which is 150 nm under the surface.  The mobility of the 2DES is
1.6$\times$$10^{5}$ cm$^{2}/$Vs at 4K, and the density is
1.02$\times$$10^{11}$ cm$^{-2}$. The 2DES under the CPW pattern
was first removed by chemical etching  before the metal-film
deposition. The 2DES left in the gap was then patterned into about
7000 identical QW mesas, each of 0.7 $\mu$m wide and 20 $\mu$m
long, by using e-beam lithography and chemical etching.  Other
experimental details can be found in our previous
publication.\cite{Hsieh2004}  Figure 11 shows a typical data,
including both the amplitude change and the phase variation as a
function of $B$. The peak power of the pulsed microwave signals
used in this measurement was $-$50 dBm (about $-$66 dBm in
average) before fed into the sample. The amplitude and phase
fluctuations in $\Delta A$ and $\Delta\phi_{s}$ data are less than
0.003\% and 0.001\textdegree, respectively. The Landau level
filling factor ($\nu$) is assigned according to the electron
density in each QW from the Shubnikov-de-Haas oscillations
observed in high-frequency data.\cite{Hsieh2004}

In summary, we have developed and demonstrated a high-sensitivity
vector detection system for very low-power microwave signals used
in a CPW broadband sensor. This system is a very powerful tool in
studying the dynamic behaviors of LDESs at low temperatures.

\section*{Acknowledgements}

This work was supported by the National Science Council of the
Republic of China under Contract No. NSC93-2112-M005-009.


\newpage

\begin{thebibliography}{99}

\bibitem{Engel1993} L. W. Engel, D. Shahar, C. Kurdak,
and D. C. Tsui, Phys. Rev. Lett. \textbf{71}, 2638 (1993).

\bibitem{Li1997} C.-C. Li, L. W. Engel, D. Shahar, D. C. Tsui,
and M. Shayegan, Phys. Rev. Lett. \textbf{79}, 1353 (1997).

\bibitem{Ye2002} P. D. Ye, L. W. Engel, D. C. Tsui, R. M. Lewis,
L. N. Pfeiffer, and K. West, Phys. Rev. Lett. \textbf{89}, 176802
(2002).

\bibitem{Lewis2002} R. M. Lewis, P. D. Ye, L. W. Engel, D. C. Tsui,
L. N. Pfeiffer, and K. W. West, Phys. Rev. Lett. \textbf{89},
136804 (2002).

\bibitem{Chen2003} Y. Chen, R. M. Lewis, L. W. Engel, D. C. Tsui,
P. D. Ye, L. N. Pfeiffer, and K. W. West, Phys. Rev. Lett. \textbf{91}, 016801 (2003).

\bibitem{Grodnensky1994} I. Grodnensky, D. Heitmann, K. v. Klitzing, K. Ploog, A. Rudenko,
and A. Kamaev, Phys. Rev. B \textbf{49}, 10778 (1994).

\bibitem{Ye2002B} P. D. Ye, L. W. Engel, D. C. Tsui, J. A.
Simmons, J. R. Wendt, G. A. Vawter, and J. L. Reno, Phys. Rev. B
\textbf{65}, 121305 (2002).

\bibitem{Lewis2001} R. M. Lewis and J. P. Carini,
Phys. Rev. B \textbf{64}, 073310(2001).

\bibitem{Hohls2001} F. Hohls, U. Zeitler, and R. J. Haug,
Phys. Rev. Lett. \textbf{86}, 5124 (2001).

\bibitem{Hsieh2004} W. H. Hsieh, Y. W. Suen, S. Y. Chang, L. C. Li,
 C. H. Kuan, B. C. Lee, and C. P. Lee, Appl. Phys. Lett. \textbf{85}, 4196 (2004).

\bibitem{White1993} R. White, \textit{Spectrum and Network
Measurements} (PTR Prentice Hall, New Jersey, 1993).

\bibitem{PLL} For the design of PLLs, please see, e.g., Roland E. Best, \textit{Phase-Locked Loops}, 5th ed. (McGraw-Hill, Inc., 2003).

\bibitem{Wixforth1989}A. Wixforth, J. Scriba, M. Wassermeier,
J. P. Kotthaus, G. Weimann, and W. Schlapp, Phys. Rev. B
\textbf{40}, 7874 (1989).

\bibitem{TTi} Thurlby Thandar Instruments Ltd., Huntingdon, Cambridgeshire PE29 7DR,  U.K.



\bibitem{lockin} SR830 from Stanford Research Systems, Inc., Sunnyvale,
CA.

\bibitem{minicircuits} Mini-Circuits, Brooklyn, NY.

\bibitem{maxim} Maxim Integrated Products, Sunnyvale, CA.

\bibitem{intersil}  Intersil Corporation,  Milpitas, CA.

\bibitem{ADI} Analog Devices, Inc.,  Norwood, MA.

\bibitem{discharge} In this design, we need to insert a TTL
invertor between the \textit{TTL}1 and the control lead of the
analog switch since there is a built-in invertor between input and
the switch control in ADG201HS.

\bibitem{vendors} MITEQ, Hauppauge, NY; Quinstar Technology Inc., Torrance, CA; M/A COM, Lowell, MA; Pulsar Microwave Corp., Clifton, NJ; Agilent, Palo Alto, CA.
\bibitem{Suen2002} Y. W. Suen, W. H. Hsieh,  L. C. Li, T. C. Wan,
 C. H. Kuan, S. D. Lin, C. P. Lee, and H. H. Cheng, \textit{Proc. of the 15th Intl. Conf. on High Magnetic Field in Semiconductor Phys.}, Oxford, U.K., B48
 (2002).

\end{thebibliography}
\end{document}